\documentstyle[aps,pre,multicol,epsf,]{revtex}
\newcommand{\be}{\begin{equation}}
\newcommand{\ee}{\end{equation}}
\begin{document}
\title{\bf The Dynamics of Money}

\author{Per Bak$^{1,2}$, Simon F. N\o rrelykke$^1$, and Martin Shubik$^2$}
\address{$^1$Niels Bohr Institute, Blegdamsvej 17, 2100 Copenhagen,
 Denmark.}
\address{$^2$Cowles Foundation for Research in Economics, Yale
University, PO Box 208281, New Haven, CT 06520, USA}
\date{\today}

\maketitle

\begin{abstract}
We present a dynamical many-body
theory of money in which the value of money is a time dependent  
``strategic variable'' that is chosen by the individual agents.
The value of money in equilibrium is not fixed by the equations, and
thus represents a continuous symmetry. The dynamics breaks this continuous 
symmetry by fixating the value of money at a level which depends on
initial conditions. The fluctuations around the equilibrium, for instance
in the presence of noise, are governed by the ``Goldstone modes''
associated with the broken symmetry. The idea is illustrated by
a simple network model of monopolistic vendors and buyers.

\end{abstract}

{PACS numbers: 02.50.Ga, Le, 05.40.-a, 05.65.+b, 05.70.Ln, 89.90.+n}

\begin{multicols}{2}

\section{Introduction}

In classical equilibrium theory in economics~\cite{Debreu}, agents submit 
their demand-vs-price functions to a ``central agent'' who then 
determines the 
relative prices of goods and their allocation to individual agents.
The absolute prices are not fixed, so the process does not determine the 
value of money, which merely enters as a fictitious quantity that
facilitates the calculation of equilibrium. Thus, traditional equilibrium 
theory does not offer a fundamental explanation 
of money, perhaps the most essential quantity in a modern economy. 

Indeed, a ``search-theoretic'' approach to monetary economics has
been proposed~\cite{KW,TW,1999}. Agents may be either money 
traders, producers,
or commodity traders. They randomly interact with each other, and
they decide whether or not to trade based on ``rational expectations''
about the value of a transaction. After a transaction the agent 
changes into one of the two other types of agents. This theory has a
steady state where money circulates.
As other equilibrium theories, this theory does not describe a dynamics
leading to the steady state, of sufficient detail for one to simulate it.

In equilibrium theory, all agents act simultaneously and globally. 
In reality, agents usually make decisions locally and sequentially. Suppose an 
agent has apples and wants oranges. He might have to sell his apples to 
another agent before he buys oranges from a third agent: hence money is needed 
for the transaction, supplying liquidity. It stores value between transactions.

Money is essentially a dynamical phenomenon, since it is intimately 
related to the temporal sequence of events. Our goal is to describe the 
dynamics of money utilizing ideas and concepts from theoretical physics
and economics, and to 
show how the dynamics may fix the value of money.

We study a network of vendors and buyers, each of whom has a simple 
optimization strategy. Whenever a transaction is considered,
the agent must decide the value of the goods and services in question, 
or, equivalently, the value of money relatively to that of the goods 
and services he intends to buy or sell. He will associate that value
to his money that he believes will maximize his utility. Thus, the 
value of money is a ``strategic variable'' that the agent in principle
is free to choose as he pleases. However, if he makes a poor choice he will
loose utility.

For simplicity, we assume that agents are rather myopic: they have short 
memories, and they take into account only the properties of their 
``neighbours'', i.e., the agents with which they interact directly. They
have no idea about what happens elsewhere in the economy.

Despite the bounded rationality of these agents, 
the economy self-organises to an equilibrium state where
there is a spatially homogeneous flow of money. 
Since we define the dynamics explicitly, we are, however, also able to treat
the nature of this relaxation to the equilibrium state, as well as 
the response of the system to perturbations and to noise-induced 
fluctuations around 
the equilibrium. These phenomena are intimately related to the
dynamics of the system, and cannot be discussed within 
any theory concerned only with the equilibrium situation.

Our model is a simple extension of Jevons'~\cite{Jevons} example of a three 
agent,
three commodity economy with the failure of the double coincidence of 
wants,
i.e., when only one member of a trading pair wants a good owned by the 
other.
A way out of the paradox of no trade where there is gain to be obtained 
by all,
is to utilize a money desired by and held by all.  Originally this was 
gold, but here
we show that the system dynamics can attach value to ``worthless'' paper 
money.

We find that the value of money is fixed by a
``bootstrap'' process: agents are forced to accept a specific value 
of money, despite this value's global indeterminacy.  
The value of money is defined by
local constraints in the network, not by trust. 
By ``local,'' we simply mean that each agent
interact only with a very small fraction of other agents in his
neighbourhood.

This situation is very similar to problems with continuous symmetry in 
physics. Consider, for instance, a lattice of interacting atoms
forming a crystal. The crystal's physical properties, including its energy, 
are not affected by a uniform translation $X$ of all atoms, 
this translational symmetry is continuous. Nevertheless, the position $x(n)$
of the $n$th atom is restricted by the position of its neighbours. This
broken continuous symmetry results in slow, large-wavelength 
fluctuations,
called {\em Goldstone modes}~\cite{Hohenberg,Mermin} or ``soft
modes.'' These modes are
easily excited thermally, or by noise, and thus gives rise to large
positional fluctuations.

\section{The Model:}
In our model, we consider $N$ agents, $n = 1,2,\ldots ,N$, placed on a
one-dimensional lattice with periodic boundary conditions. This geometry
is chosen in order to have a simple and specific way of defining who is 
interacting
with whom. The geometry is not important for our general conclusions
concerning the principles behind the fixation of prices.

We assume that agents cannot consume their own output, so in order to consume
they have to trade, and in order to trade they need to produce.
Each agent  produces a quantity $q_{n}$, of one good, which is sold
at a unit price $p_n$, to his left neighbour $n-1$. 
He next buys and 
consumes one good from his neighbour to the right, who  
subsequently buys the
good of {\em his\/} right neighbour, etc., until 
all agents have made two transactions. This process 
is repeated indefinitely, say, once per day.

For simplicity, all agents are given utility
functions of the same form

\be
 u_n = -c(q_{n}) + d(q_{n+1}) + 
	     I_{n}\! \cdot \! (p_{n}q_{n}- p_{n+1}q_{n+1}) \enspace.
  \label{eq:utility}
\ee

The first term, $-c$, represents the agent's cost, or displeasure,
associated with producing $q_{n}$ units of the good he produces. 
The displeasure is an increasing 
function of $q$, and $c$ is convex, say, because the agent gets 
tired. The second term $d$, is his utility of the good he 
can obtain from his neighbour.
Its marginal utility is decreasing with $q$, 
so $d$ is concave.
This choice of $c$ and $d$ is common in economics;
see, e.g., \cite{TW}.

An explicit example 
is chosen for illustration and analysis,

\begin{equation}
  c(q_n) =  a q_n^\alpha  \mbox{\hspace{0.5em},\hspace{1.5em}} 
  d(q_{n+1}) = b q_{n+1}^\beta \enspace .
\end{equation}
The specific values of $a$, $b$, $\alpha$, and $\beta$ are
not important for the general results, as long as $c$ remains 
convex and $d$ concave.
For our analysis we choose $a=\frac{1}{2}$, $b=2$, $\alpha=2$,
and $\beta=\frac{1}{2}$.

The last term represents the change in utility associated with the gain
or loss of money after the two trades.
Notice that the dimension of $I_n$
is [utility per unit of currency], i.e., the physical interpretation
is the {\em value of money.} 

Each agent has knowledge only about the utility functions 
of his two
neighbours, as they appeared the day before.
The agents are monopolistic, i.e., agent $n$ sets the
price of his good, and agent $n-1$ then decide how much $q_{n}$, he 
will buy
at that price. This amount is then produced and sold---there is no 
excess
production. The goal of each agent is to maximize his utility, 
by adjusting $p_n$ and $q_{n+1}$, while
maintaining a constant (small) amount of money. Money has value only as 
liquidity. There is no point in keeping money, all that is needed is what
it takes to complete the transactions of the day.

Thus, the agents aim to achieve a situation where the expenditures are 
balanced by
the income:

\be
  p_{n}q_{n} -  p_{n+1}q_{n+1} = 0 \enspace .
  \label{eq:constraint}
\ee
 
When the value of money is fixed, $I_n$ =$I$, the agents
optimize their utility by charging a price

\be
  p = 2^{\frac{1}{3}} \cdot I^{-1}
 \label{eq:moneq1}
\ee
and selling an amount
\be
  q =  2^{-\frac{2}{3}}
  \label{eq:moneq2}	
\ee
at that price. This is the monopolistic equilibrium.

Note that the resulting quantities $q$, are independent
of the value of money, which thus represents a continuous
symmetry. There is nothing in the equations that fixes the
value of money and the prices. Mathematically, the continuous symmetry
expresses the fact that the equations for the quantities are
``homogeneous of order one.'' The number of equations is one
less than the number of unknowns, leaving the value of money
undetermined. We shall see how this continuous symmetry
eventually is broken by the dynamics.

Agent $n$ tries to achieve his goal by estimating
the amount of goods $q_n$, that his neighbour will 
order at a given price, 
and the price $p_{n+1}$, that his other neighbour will charge at the
subsequent transaction.

Knowing that his
neighbours are rational beings like himself, he is able to 
deduce the functional relationship between the price $p_n$,
that he demands and the amount of goods $q_n$, that will be 
ordered in response to this.
Furthermore, he is able to estimate the 
size of $p_{n+1}$, based on the previous transaction with
his right neighbour. This enables him to decide what the 
perceived value of money should be, and hence how much
he should buy and what his price should be.
This process is then continued indefinitely,
at times $\tau = 1,2,3, \ldots$.

This defines the game. The strategy we investigate contains the 
assumption that agents do not change their valuation of money
$I$, between their two daily transactions, and they maximize their
utility accordingly.

The process is initiated by choosing some initial
values for
the $I$'s. They could, e.g., be related to some former gold 
standard.

In fixing his price at his first transaction of day $\tau$, agent $n$ 
exploits the knowledge he has
of his neighbours' utility functions, i.e., he knows that the agent to 
the left will maximize his function with respect to $q_{n,\tau}$

\be
  \frac{\partial u_{n-1,\tau}}{\partial q_{n,\tau}} =  0 \enspace;
\ee 
hence the left neighbour will order the amount:
\be
  q_{n,\tau}  =  \left( I_{n-1,\tau} \, p_{n,\tau} \right) ^{-2}\enspace .
\label{eq:max1}
\ee

This functional relationship between the amount
of goods $q_{n,\tau}$, ordered by agent $n-1$ at time $\tau$
and the price $p_{n,\tau}$, set by agent $n$, allows agent $n$
to gauge the effect of his price policy. Lacking knowledge about
the value of $I_{n-1,\tau}$, agent $n$ instead estimates it to equal
the value it had in the previous transaction $I_{n-1,\tau-1}$,
which he knows.
Eliminating $q_{n,\tau}$ from Eq.~(\ref{eq:utility}) we obtain

\begin{eqnarray}
 u_{n,\tau} &=& -\frac{1}{2}I_{n-1,\tau-1}^{-4}p_{n,\tau}^{-4} 
  + 2\sqrt{q_{n+1,\tau}} 
  \label{eq:specific} \\
  & & \mbox{} + I_{n,\tau}\! \cdot \! 
  (p_{n,\tau}^{-1}I_{n-1,\tau-1}^{-2} - p_{n+1,\tau}q_{n+1,\tau}) 
  \nonumber \enspace .
\end{eqnarray}
Maximizing this utility $u_{n,\tau}$, with respect to 
$p_{n,\tau}$ and $q_{n+1,\tau}$ yields

\begin{equation}
  p_{n,\tau}  =
  2^{\frac{1}{3}}I_{n,\tau}^{-\frac{1}{3}}
  I_{n-1,\tau-1}^{-\frac{2}{3}}\enspace,
  \label{eq:pn}
\end{equation}
and
\be
  q_{n+1,\tau} = \left( I_{n,\tau} p_{n+1,\tau} \right) ^{-2} \enspace .
 \label{eq:qn+1}
\ee
By arguments of symmetry,

\begin{equation}
  p_{n+1,\tau}  =
  2^{\frac{1}{3}}I_{n+1,\tau}^{-\frac{1}{3}}
  I_{n,\tau-1}^{-\frac{2}{3}}
  \label{eq:pn+1}
\end{equation}
is the price agent $n+1$ will demand of agent $n$ in the 
second transaction.
Since agent $n$ does not yet know the value of $I_{n+1,\tau}$, 
he instead uses the known value of $I_{n+1,\tau-1}$ when
estimating $p_{n+1,\tau}$.

In the constraint Eq.~(\ref{eq:constraint}), the following
expressions are used
\begin{eqnarray}
  q_n &=& q^{\mbox{\scriptsize (guess)}}_{n,\tau} =
    \left( I_{n-1,\tau-1} p_{n,\tau} \right) ^{-2} \enspace,\\
  p_{n+1} &=& p^{\mbox{\scriptsize (guess)}}_{n+1,\tau} =
  2^{\frac{1}{3}} I_{n+1,\tau-1}^{-\frac{1}{3}}
  I_{n,\tau -1}^{-\frac{2}{3}} \enspace,\\
  \label{eq:pn+1guess}
  q_{n+1} &=& q^{\mbox{\scriptsize (guess)}}_{n+1,\tau}=
    \left( I_{n,\tau} 
  p^{\mbox{\scriptsize (guess)}}_{n+1,\tau} \right) ^{-2} \enspace,
\end{eqnarray}
and $p_n$ is given by Eq.~(\ref{eq:pn}). 
Solving for $I_{n,\tau}$, and evaluating at time $\tau+1$,
we find~\cite{math}

\begin{equation}
  I_{n,\tau+1} = \left( I_{n-1,\tau }^4 \, I_{n,\tau }^2 \,
  I_{n+1,\tau }\right) ^{\frac{1}{7}}
 \label{eq:In}
\end{equation}
which sets agent $n$'s value of money on day $\tau+1$
equal to a weighted geometric average of the value 
agent $n$ and his two neighbours
prescribed to their money the previous day. 
Using this value of $I_{n}$, 
agent $n$ can fix his price $p_{n}$ and decide 
which quantity $q_{n+1}$, he should optimally buy.
This simple equation completely specifies the dynamics of
our model. The entire strategy can be reduced to an update scheme
involving only the value of money---everything else follows from this.
Thus, the value of money can be considered the basic strategic variable.

Although Eq.~(\ref{eq:In}), has been
derived for a specific simple example, we submit that the
structure is much more general. In order to 
optimize his utility function, the agent is forced to
accept a value of money, and hence prices, which pertain
to his economic neighbourhood. Referring again to a
situation from physics, the position of an atom on a general lattice
is restricted by the positions of its neighbours, despite the
fact that the entire lattice can be shifted with no 
physical consequences.

Even though there is no utility in the possession of money, 
as explicitly expressed by Eq.~(\ref{eq:constraint}), 
the strategies and dynamics of the model nevertheless leads to a 
value being ascribed to the money. 
The dynamics in this model is driven by the need of
the agents to make estimates about the coming transactions. 
In a sense, this models the
real world where agents are forced to make plans about the
future, based on knowledge about the past---and, in practise,
only a very limited part of the past.
In short: the dynamics is generated by the bounded 
rationality of the agents.

In the steady state, where the homogeneity of the utility
functions give $I_n = I_{n+1}$, we retrieve the monopolistic
equilibrium equations~(\ref{eq:moneq1}) and (\ref{eq:moneq2}).

\section{Solving the dynamics:}
Taking the logarithm in Eq.~(\ref{eq:In}) and introducing 
$h_{n,\tau }= \ln (I_{n,\tau })$ 
yields the linear equation:

\begin{equation}
  h_{n,\tau +1}= \mbox{\small $\frac{4}{7}$}h_{n-1,\tau } +  
  \mbox{\small$\frac{2}{7}$}h_{n,\tau}
  +  \mbox{\small$\frac{1}{7}$}h_{n+1,\tau }\enspace ,
 \label{eq:log}
\end{equation}
describing a Markov process.
Now assume that $h_{n,\tau}$ is a slowly varying function of $(n,\tau)$
and that we may think of it as the value of a differentiable function
$h(x,t)$ in $(x,t)=(n\delta x,\tau \delta t)$. Then, expanding to 
first order in $\delta t$ and second order in $\delta x$, 
we find the diffusion equation

\begin{equation}
  \frac{\partial h(x,t)}{\partial t}= D \frac{\partial ^2 
  h(x,t)}{\partial x^2}
  - v \frac{\partial h(x,t)}{\partial x} \enspace , 
\label{eq:diffus}
\end{equation}
with diffusion coefficient 
$D = \frac{5}{14}\frac{(\delta x)^2}{\delta t}$, 
and convection velocity $v = \frac{3}{7}
\frac{\delta x}{\delta t}$. 
The generator $T$, of infinitesimal time translations is defined by

\begin{equation}
  \frac{\partial h(x,t)}{\partial t}= T h(x,t)
 \enspace .
\end{equation}

Taking the lattice Fourier 
transformation, the eigenvalues of $T$ are found
to be $\lambda_k = -k^2 D - i k v$,
where the periodic boundary condition 
yields $k=\frac{2\pi}{N}\, l$; $l=0,1,\ldots,N-1$.
The damping time for each mode $k$,
is given by  $t_k = (k^2 D)^{-1}$, i.e., it increases
as the square of the system size $N$.
The only mode that is not dampened has
$k=0$, and is the soft
``Goldstone mode''~\cite{Hohenberg,Mermin} associated with the 
broken continuous symmetry  with respect to a uniform shift 
of the logarithm of prices in the equilibrium:

All prices can be changed by a common factor, but the amount of
goods traded will remain the same.
The rest of the modes are all dampened (for a finite-size system), 
and hence the system eventually relaxes to the steady state.

\begin{figure}
\input{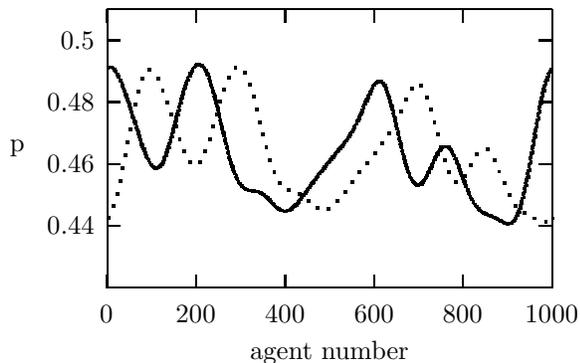}
\begin{minipage}[]{8cm}
\caption[]{\label{fig1}Variation of prices for all agents at two 
	different times, $\tau = 3000$ (full line) and  
	$\tau=3200$ (broken line.)}
\end{minipage}
\end{figure}

\begin{figure}
\input{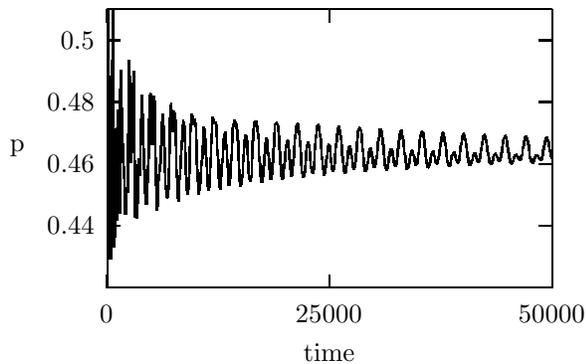}
\begin{minipage}[t]{8cm}
\caption[]{\label{fig2}Price variation for a single agent. 
	The oscillations are an artifact
	due to the periodic boundary condition,
	setting $h_{N+1,\tau} = h_{1,\tau}$.} 
\end{minipage}
\end{figure}

Figures~\ref{fig1} and \ref{fig2} show results from a 
numerical solution 
of Eq.~(\ref{eq:log}) for 1000 agents
with random initial values for the variables $h$
(sampled from a uniform distribution on the interval [0,2].) 
Figure~\ref{fig1} shows the spatial variation of prices
at two different times---convection with velocity
$v = \frac{3}{7}\frac{\delta x}{\delta t}$ is clearly seen, 
while
the effect of diffusion is not visible on this time scale. 
The relatively weak effect of diffusion means that spatial 
price variations, such as those shown in Fig.~\ref{fig1},
can travel around the entire lattice many times before 
diffusion has evened them out.
Consequently, the individual agent
experiences price oscillations with slowly decreasing amplitude, 
as seen in Fig.~\ref{fig2}.

Thus, despite the myopic behaviour of agents, the system
evolves towards an equilibrium. 
But in contrast to equilibrium theory, we obtain
the temporal relaxation rates towards the equilibrium, as well as 
specific absolute values for individual prices. The value of
money is fixed by the history of the dynamical process, i.e.,
by the initial condition combined with the actual strategies by
the bounded rational agents.

\section{Noise}

If an agent is suddenly supplied with some extra amount of money,
he will lower his value of money, hence increase his price 
and consequently work less and buy more goods,
the effect being inflation propagating through the system, as
described by the solution to Eq.~(\ref{eq:diffus}) for a
delta-function initial condition~\cite{solution}. 
Likewise, the destruction or loss of some amount of money 
by a single agent will affect the whole system.
These are both transient effects, and in the steady state
the same amount of goods will be produced and consumed,
as before the change.

In general, there might be some noise in the system, due to 
imperfections
in the agents' abilities to optimize properly their utility 
functions, or due to
external sources affecting the utility functions.
A random multiplicative error
in estimating the value of money transforms to a linear noise in 
Eq.~(\ref{eq:diffus}).
We assume that the noise $\eta(x,t)$, has the characteristics:
$\langle \eta (x,t) \rangle = 0$ and $\langle \eta(x,t) \eta(x',t') 
\rangle = A \delta (x-x') \delta(t-t')$. 
Adding it to Eq.~(\ref{eq:diffus})
and taking the Fourier transform (with periodic boundary conditions
in a system of size $L$)
one finds the equal-time correlation function:

\begin{eqnarray}
	\lefteqn{\langle [h(x) - h(0)] 
	[h(x') -   h(0)] \rangle} \nonumber \ \ \ \\
   & = &  \frac{A}{2DL} \sum_q q^{-2} (e^{iqx}-1)(e^{-iqx'}-1) \enspace,
\end{eqnarray} 
where $q = \frac{2\pi}{L}n;\mbox{ } n=0,\pm 1, \pm 2,\ldots$.
For $x = x'$ and $L\rightarrow \infty $ this becomes

\be
	\langle [h(x) - h(0)]^2 \rangle
	= \frac{A}{2D} x \enspace,
\ee
viz., the dispersion for a biased random walker in one
dimension with position $h$, time $x$, and diffusion coefficient
$\frac{A}{4D}$.
In the presence of noise, the agents no longer agree about the 
value of
money, and there will be large price fluctuations. The fluctuations
reflect the lack of global restoring force due to the continuous
global symmetry.

How much money is needed to run an economy? In this 
model-economy the total amount of money is reflected 
in the agents' $I$'s, and is always conserved, 
as seen by
\be
  \sum_n (p_{n}q_{n} - p_{n+1}q_{n+1}) = 0\enspace,
\ee
since we have periodic boundary conditions.
No matter what the initial amount of money
in the system is, the system will go to the equilibrium where precisely 
that amount is needed---the final $I$'s are fixed by the initial money
supply. The total amount of money in the economy is irrelevant, since the 
utility
and amount of goods exchanged in the final equilibrium does not depend on
that. However, as previously described,
changes in the amount of money have interesting transient effects.

\section{Conclusion:}
Here we considered a simple toy model with simple monopolistic agents. 
In general, economy deals with complicated heterogeneous
networks of agents, with complicated links to one another, representing
the particular ``games'' they play with one another.
We submit
that the general picture remains the same. At each trade, the agents 
evaluate
the value of money, by analysing their particular local situation,
and act accordingly. The prices charged by the 
agents will
be constrained by those of the interacting agents. It would be 
interesting
to study the formation and stability of markets where very many 
distributed
players are interested in the same goods, but not generally interacting 
directly with
one another.  Indeed, we have considered an allied model with a market 
structure introduced    
in a related, more explicitly economics-oriented discussion~\cite{1999}.

Modifications of this network model may also provide a toy laboratory for the 
study of the effects of
the introduction of the key financial features of credit and bankruptcy 
as well as the control
problems posed by the governmental role in varying the money supply.

\end{multicols}


\begin{references}
\bibitem{math} This is mathematically equivalent to differentiating
  Eq.~(\ref{eq:specific}) with respect to $I_{n,\tau}$ 
	(with the expressions in 
  	Eqs.~(\ref{eq:pn},\ref{eq:qn+1},\ref{eq:pn+1}) 
	substituted for $p_{n,\tau}$, 
	  $q_{n+1,\tau}$, and $p_{n+1,\tau}$) and demanding it be zero.
\bibitem{solution} The solution to Eq.~(\ref{eq:diffus}) for a 
	delta-function initial condition is: 
	$h(x,t) = \frac{C}
	{\sqrt{4 \pi D t}}\sum_{j=-\infty}^{\infty} \exp \left(- 
	\frac{(x - jL - vt)^2}{4 D t}\right)$, 
	where $L$ is the system size and $x\in[0,L]$.
\end{references}
\end{document}